\documentstyle[hx37_latex]{article}

\input{psfig}
\def\be{\begin{equation}}
\def\ee{\end{equation}}
\def\bea{\begin{eqnarray}}
\def\eea{\end{eqnarray}}
\begin{document}

\title{GRAVITATIONAL LENSING AND THE REDSHIFT DISTRIBUTION OF B$>25$
GALAXIES}

\author{ Y. MELLIER }

\address{Institut d'Astrophysique de Paris, 98bis Boulevard Arago \\ 
75014 Paris, France}


\maketitle\abstracts{
In addition to the determination of the mass distribution of lenses
 (see P. Schneider, this conference), 
strong and weak lensing can also be powerfull tools to analyse the
redshift distribution of faint galaxies.  In this review, I summarize 
the present status of redshifts of galaxies beyond $B=25$ as they are
inferred by spectroscopy of magnified galaxies, lensing inversion and 
magnification bias.}

\section{Introduction}

During the last decade, outstanding results have been obtained on the 
redshift distribution of faint galaxies up to 
 $I=22.5$  (see O. Le F\`evre, this conference) or $I=23$ (Cowie et al. 1996). 
With the coming of 10 meter 
class telescopes equipped with wide field multi-object
spectrographs these surveys will be extended to thousands of galaxies.
 The leading goals of these
programme is to explore the evolution of clustering of galaxies, 
their physical and stellar evolution with redshift up to 
$B=24-25$ or $I=22.5-23.5$ as well as to study very high redshift galaxies. 
The galaxies with magnitudes $B>25$ are also important for the models of
galaxy formation since we do not know whether they are all at large
redshift or if there is a significant fraction of faint nearby dwarfs
galaxies.  Furthermore, the weak lensing inversion uses 
the grid of  faint distant sources with 
magnitudes between $B=25$ to $B=28$ for which the redshift distribution
is unknown. While this information 
 is not important for nearby lenses,  it is crucial for 
 for those having redshifts larger than $z=0.5$ and can be a major source of 
uncertainty in the mass
determination (see Luppino \& Kaiser 1996). In the perspective of the  
new surveys to study the large-scale mass distribution by using
 weak lensing, the need of the redshift distribution of 
the faintest galaxies is thus esssential.

Unfortunately, while we do expect considerable informations in the magnitude 
range $B<25$,
beyond this limit even 10 meter class telescopes are still too small and
redshifts of a complete sample of $B>25$ galaxies cannot
be secured in a reasonable amount of observing time. The possibility of using 
photometric
redshifts has been proposed as early as the beginning of eighties
 and has been  studied in great details. Observations 
as well as reliability tests are still underway (Pell\'o 
private communication). Howewer, they are based on
theoretical evolution scenarios of galaxies whose predictions about
faint distant galaxies are not confirmed yet. Furthermore, there is no 
hope to calibrate the photometric redshifts of the faint samples
 with spectroscopic data. 

The most attractive alternative is the use of the  deviation,  magnification 
and distortion effects induced by gravitational lensing on extended objects. 
 In this review, I discuss the recent advances in the redshift
distribution of $B>25$ galaxies by using spectroscopic samples of
arc(let)s, the lensing inversion and finally the magnification bias.
Though many works in these fields are still underway, they are new
promising approaches that must be tested jointly with photometric
redshifts in order to cross-check their reliablity and the consistency 
of their predictions.
 
\begin{figure}
\rule{5cm}{0.2mm}\hfill\rule{5cm}{0.2mm}
%\vskip 2.5cm
%\rule{5cm}{0.2mm}\hfill\rule{5cm}{0.2mm}
\hskip 1.0 cm
\psfig{figure=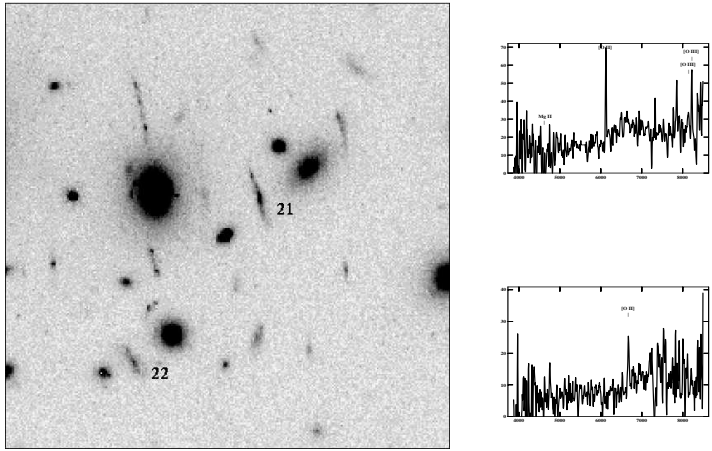,height=6. cm}
\caption{Redshift of faint arclets in lensing-clusters. The left panel is a
deep HST image of A2390.  Arclets are visible with
some of them showing multiple images with image parity changes. The
arclets 21 and 22 have been observed by B\'ezecourt and Soucail (1996). 
As expected, they show the
[OII]$\lambda$3727 emission line from which the redshifts are easily
measured   ($z=0.643$ and $z=0.790$ respectively. Courtesy J.
B\'ezecourt)..
\label{fig:radish}}
\end{figure}

\section{Spectroscopic surveys of arc(let)s}
Spectroscopic redshifts of arc(let)s are indispensable  
to calculate the angular distances $D_{\rm d},
D_{\rm ds}$ and $D_{\rm s}$ and to get the absolute scaling of the projected
mass density. The redshifts of a large number of arc(let)s 
in each
individual lensing-clusters provide the positions of many critical lines
and allow to probe locally the mass distribution. In practice, the
redshifts of arc(let)s check the lens modelling obtained
from giant arcs and can be used to refined it. It 
is also possible to obtain information on the cosmological parameters if one 
could have enough redshift to constrain both the deflector and the geometry of
the Universe. 

More recently, with the development of the lensing inversion technique 
(see next section), the need of spectroscopic confirmations of its 
predictions led to intensive observations of arclets. 
Spectroscopic surveys of the "brightest"
arclets in many clusters  are now underway and 
first results have been obtained in A2390 (B\'ezecourt \& Soucail 1996)
and A2218 (Ebbels et al. 1996; Ebbels this conference). 
 Since most of these objects are very faint, only arclets showing bright spots 
of stars forming regions on HST images are selected in order to detect
an emission line and to optimize the chance to get reliable redshift 
(see figure 1). 

About 20 redshifts of arcs and 20 redshifts of arclets have been
measured. However, the use of this sample to recover the redshift 
distribution of
 $B>25$ galaxies is difficult because it is biased. First, 
 only arclets with star forming emisson lines are
selected. Second, beyond $B=25$ the
magnification bias  favours observations of blue galaxies rather
than red.
So, even if the spectroscopy of arclets is 
crucial for the lens modelling and eventually to obtain the spectral
energy distribution of high-redshift galaxies, the redshift
distributions obtained from these methods are still questionnable.

\begin{figure}
\rule{5cm}{0.2mm}\hfill\rule{5cm}{0.2mm}
%\vskip 2.5cm
%\rule{5cm}{0.2mm}\hfill\rule{5cm}{0.2mm}
\psfig{figure=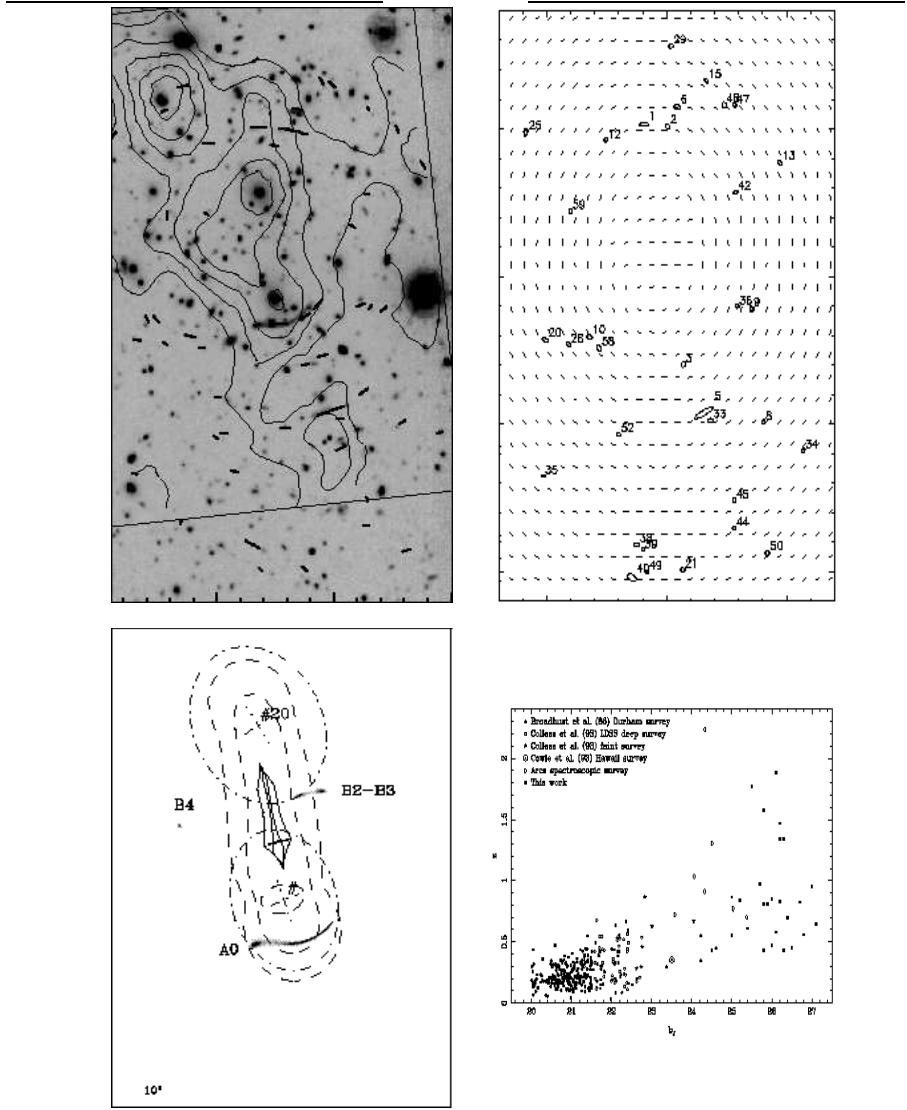,height=14.7 cm}
\caption{The top
left pannel shows a B image of A370 with the ROSAT isocontours and the 
position of arclets superimposed (dark segments).  They are consistent 
with the lens modelling (bottom left) and we can 
expect good redshift predictions of arclets from the reconstructed
 shear map (top right). Kneib et
al. inverted the arclets in A370 and found the magnitude-redshift
diagramme plotted in the bottom right panel. However, the ROSAT 
isophotes and the arclets positions show some discrepancies with the
modelling on the eastern side (left) and in this region the predicted
redshifts could be wrong. 
\label{fig:radish}}
\end{figure}

\section{Redshifts of arclets from lensing inversion}
When it is possible to recover the lensing potential with a good
accuracy, the lensing equation can be  inverted to  
reconstruct the lensed images back to the source plane.  This is basically 
the procedure of the lensing inversion which searches 
for each arclet the  source plane where its 
distortion is minimum, assuming it gives its most 
 {\sl probable redshift}.  The obvious interest of this 
method is that it does not depend on the magnitude of the arclet but on
its position and its shape in the image plane. Potentially, it provides
redshift of galaxies up to $B=27$.

The lensing inversion has been developped by the Toulouse/Paris group 
and was first applied on A370 (Kneib et al. 1994) from the lens
modelling of the giant arc. Though the (intrinsic)magnitude-redshift found 
for these arclets shows a good continuity with the faint spectroscopic
surveys, there are still some uncertainties. 
 In fact, as it is shown in figure 2, the X-ray isophotes and
the arclet positions do not follow the expectations of the lens
modelling in the eastern region. This is an indication that 
while the modelling is excellent in the cluster center, the mass
distribution has not a simple geometry beyond the giant arc.
Furthermore, the lensing inversion is also sensible to the accuracy of
the shape measurements of each arclet, and for so faint objects errors can
be large.

\begin{figure}
\rule{5cm}{0.2mm}\hfill\rule{5cm}{0.2mm}
%\vskip 2.5cm
%\rule{5cm}{0.2mm}\hfill\rule{5cm}{0.2mm}
\psfig{figure=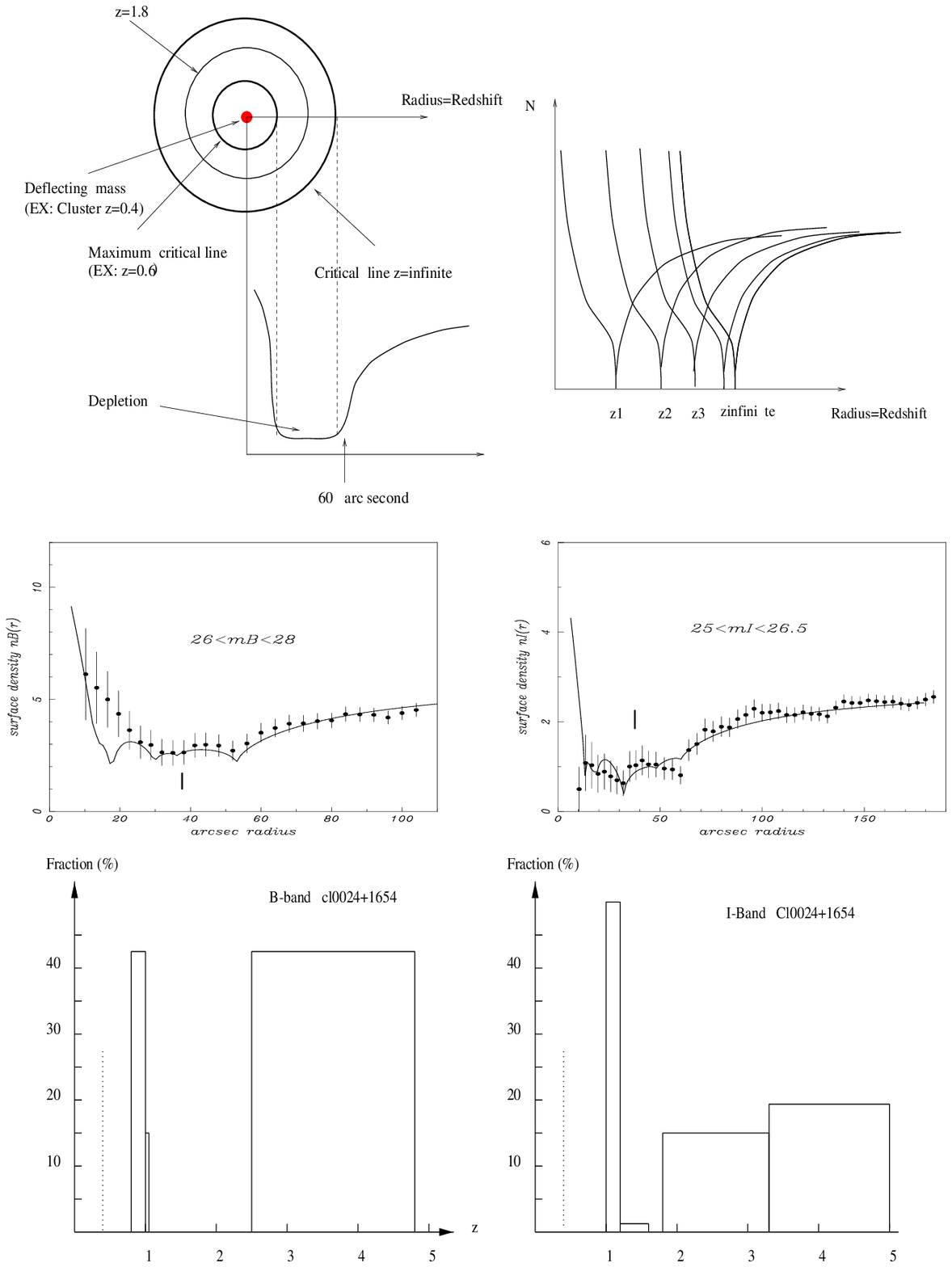,height=13.7 cm}
\caption{Depletion by a singular isothermal sphere
as it would be observed on the sky
 and radial density of galaxies (top left).
For a given redshift,
the minimum of the depletion is sharp and its radial position is
equivalent to a redshift (top right). The miminum increases with the
redshift of sources but the depletion curves tighten and converge
towards the curve corresponding to sources at infinity.  In a realistic
case, the redshift distribution is broad and the individual curves must
be added. In this case, instead of
the single peaked depletion we expect a more pronounced minimum  between
two radii (i.e. two redshifts; top left). The middle panels show the 
depletion curves observed in $B$ and $I$ in Cl0024.
 Since  the mass distribution of this lens is well known, one can recover the
redhsift of the sources for the $B$ and $I$ populations (bottom panels:
note that this is a fraction of galaxies. The width of boxes is
the redshift range, not a total number of galaxies).
\label{fig:radish}}
\end{figure}
There are two solutions to solve these issues: first, it is highly
preferable to use HST images instead of ground based images. The results
obtained by Kneib et al. (1996) on A2218 show the efficiency of the
lensing inversion when applied on excellent images. Second, it is 
important to have lensing-clusters with simple geometry. In this
respect, though A370 and A2218 are rather well modelled, they are not 
the simplest and clusters like MS0440 or
MS2137-23 are  better candidates.
 
\section{The distribution of faint galaxies from the
magnification bias }
The projected number density of galaxies through a lens   results from 
 the competition between the gravitational magnification that increases
the detection of individual objects and the deviation
of light beam that increases the area and thus decreases the apparent number 
density.  Therefore the amplitude of the magnification bias
depends on the slope of the galaxy counts as a function of magnitude and
on the magnification factor of the lens (Broadhurst et al. 1995): 
when the slope is higher than $0.4$ the number density increases,
whereas below $0.4$ is decreases and the radial distribution shows a
typical depletion curve (see figure 3).

When the slope is lower than $0.3$, a sharp decrease of the number of galaxies
is expected 
 close to the critical radius of the lens corresponding to the redshift
of the background sources. For a broad redshift distribution, 
 it can result a shallower depletion between the smallest and
the largest critical line
which depends on the redshift  distribution of the galaxies
(Figure 3).  Therefore, the analysis of the shape of the depletion
curves provide a new way
to sort out  their redshift
distribution. As the lensing inversion, this is a statistical 
 method which can also
 infer redshift of very faint sources (up to $B=28$) but does not need 
anymore information on the shapes of arclets. However, the need of a good 
lens modelling is still necessary.

 This method was first used by Fort et al (1996) in the cluster
Cl0024+1654 to study the
faint distant galaxies population in the extreme range of magnitude
$B=26.5-28$ and $I=25-26.5$. For these selected bins of magnitude they
found on their CFHT blank fields
that the counts slope was near 0.2, well suited for the study of
the  effect. After analysis of the shape of the depletion
curve (figure 4), $60\% \pm 10\%$ of the
$B$-selected galaxies were found
between $z=0.9$ and $z=1.1$ while most of the remaining $40\% \pm 10\%$
galaxies appears to be broadly distributed around a redshift of
$z=3$. The
$I$ selected population present a similar distribution  with two maxima,
but
spread up to a larger redshift range  with about 20\% above  $z > 4$ 
(figure 3).

This very first tentative must be pursued on many lensing clusters in order to
provide significant results on the redshift distribution of the faintest 
distant galaxies. Though it is a very promising approach, it also need to be
applied on clusters with simple geometry. Furthermore, the detection
procedure demands ultra-deep exposures with subarcsecond seeing.

\section{Conclusions }
The redshift distribution of galaxies beyond $B=25$ is a crucial
scientific question for galaxy evolution and weak lensing inversions. 
 I have discussed three innovative ways which can go as faint
as $B=28$.  They must be considered jointly with photometric redshifts
which will need confirmations from others observations.  But whatever 
the method, how and when will we be sure that these redshifts are correct 
from spectroscopic data? This 
key issue may be solved with ultra-deep CCD spectroscopic exposures with
the VLTs.  This is a major challenge for the coming years that will be 
possible to match with the systematic  use of gravitational telescopes.

\section*{Acknowledgments}
I thank B. Fort, R. Ellis, R. Pell\'o, P. Schneider and L. Van Waerbeke for 
  stimulating discussions on lensing and on distances of faint galaxies. 

\begin{figure}
\rule{5cm}{0.2mm}\hfill\rule{5cm}{0.2mm}
%\vskip 2.5cm
%\rule{5cm}{0.2mm}\hfill\rule{5cm}{0.2mm}
\psfig{figure=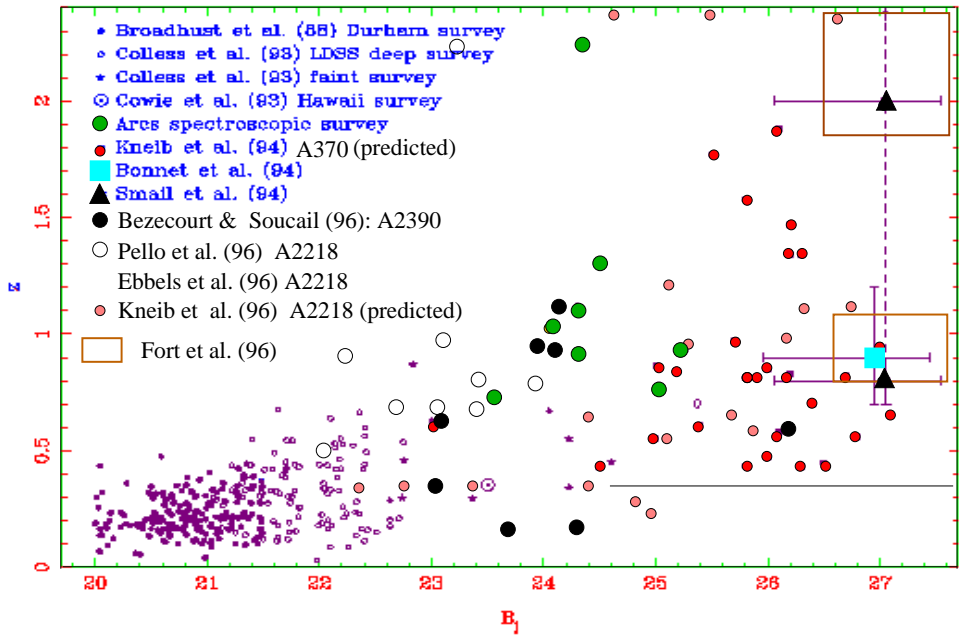,height=9.0 cm}
\caption{A magnitude-redshift diagramme showing the positions of
the redshift surveys (dark symbols on the left), the arc(let)s 
spectroscopic surveys (large circles. Those concerning A2218 
have been kindly provided by Pell\'o prior to publication), the predictions 
of lensing inversions for A370 and A2218 (small circles),   
of weak lensing studies  by Bonnet et al. and Smail et al. (triangles) 
and finally, of the depletion curves in Cl0024 (large boxes). The 
spectroscopic redshift of Cowie et al. (1996) with Keck would be between 
$B=22.5$ and $B=24.5$. We see the potential interest of gravitational
lensing which provide redshifts up to $B=28$.  The straight line on
bottom right is the redshift of A370 which is a limit of the lensing
inversion in this cluster.
\label{fig:radish}}
\end{figure}
\section*{References}
%
% The following journals are predefined in the .sty file:
%
% \def\aj{{AJ}}			
% \def\araa{{ARA\&A}}		
% \def\apj{{ApJ}}			
% \def\apjl{{ApJ}}		
% \def\apjs{{ApJS}}		
% \def\ao{{Appl.~Opt.}}		
% \def\apss{{Ap\&SS}}		
% \def\aap{{A\&A}}		
% \def\aapr{{A\&A~Rev.}}		
% \def\aaps{{A\&AS}}		
% \def\azh{{AZh}}			
% \def\baas{{BAAS}}		
% \def\jrasc{{JRASC}}		
% \def\memras{{MmRAS}}		
% \def\mnras{{MNRAS}}		
% \def\pra{{Phys.~Rev.~A}}	
% \def\prb{{Phys.~Rev.~B}}	
% \def\prc{{Phys.~Rev.~C}}	
% \def\prd{{Phys.~Rev.~D}}	
% \def\pre{{Phys.~Rev.~E}}	
% \def\prl{{Phys.~Rev.~Lett.}}	
% \def\pasp{{PASP}}		
% \def\pasj{{PASJ}}		
% \def\qjras{{QJRAS}}		
% \def\skytel{{S\&T}}		
% \def\solphys{{Sol.~Phys.}}	
% \def\sovast{{Soviet~Ast.}}	
% \def\ssr{{Space~Sci.~Rev.}}	
% \def\zap{{ZAp}}			
% \def\nat{{Nature}}		
% \def\iaucirc{{IAU~Circ.}}
% \def\aplett{{Astrophys.~Lett.}}
% \def\apspr{{Astrophys.~Space~Phys.~Res.}}
% \def\bain{{Bull.~Astron.~Inst.~Netherlands}}
% \def\fcp{{Fund.~Cosmic~Phys.}}
% \def\gca{{Geochim.~Cosmochim.~Acta}}
% \def\grl{{Geophys.~Res.~Lett.}}
% \def\jcp{{J.~Chem.~Phys.}}	
% \def\jgr{{J.~Geophys.~Res.}}	
% \def\jqsrt{{J.~Quant.~Spec.~Radiat.~Transf.}}
% \def\memsai{{Mem.~Soc.~Astron.~Italiana}}
% \def\nphysa{{Nucl.~Phys.~A}}
% \def\physrep{{Phys.~Rep.}}
% \def\physscr{{Phys.~Scr}}
% \def\planss{{Planet.~Space~Sci.}}
% \def\procspie{{Proc.~SPIE}}
%

\end{document}